\documentclass[12pt]{article}

\usepackage{amsmath,amssymb,amsfonts,amsbsy}
\usepackage{cite}
\usepackage{graphicx}
\usepackage{wrapfig}
\newcommand{\Li}{\mbox{Li}_2}
\newcommand\ba{\begin{eqnarray}}
\newcommand\ea{\end{eqnarray}}
\newcommand\be{\begin{equation}}
\newcommand\ee{\end{equation}}
\newcommand\nn{\nonumber}

\begin{document}
\addcontentsline{toc}{subsection}{{Charge asymmetry and symmetry properties}\\
{\it Egle Tomasi-Gustafsson}}

\setcounter{section}{0}
\setcounter{subsection}{0}
\setcounter{equation}{0}
\setcounter{figure}{0}
\setcounter{footnote}{0}
\setcounter{table}{0}

\begin{center}
\textbf{Charge asymmetry and symmetry properties}

\vspace{5mm}

E.~Tomasi-Gustafsson
\vspace{5mm}

\begin{small}
   \emph{CEA,IRFU,SPhN, Saclay, 91191 Gif-sur-Yvette Cedex, France, and } \\
  \emph{CNRS/IN2P3, Institut de Physique Nucl\'eaire, UMR 8608, 91405 Orsay, France} \\
 \emph{E-mail: etomasi@cea.fr}
\end{small}
\end{center}

\vspace{0.0mm} 

\begin{abstract}
Applying general symmetry properties of electromagnetic interaction, information from electron proton elastic scattering data can be related to charge asymmetry in the annihilation channels $e^++e^-\leftrightarrow \bar p + p$ and to the ratio of the cross section of elastic electron and positron scattering on the proton. A compared analysis of the existing data allows to draw conclusions on the reaction mechanism. 

\end{abstract}

\vspace{7.2mm} 

Elastic and inelastic electron scattering has been considered the most direct way to learn about the internal structure of hadrons. If one assumes that the  underlying mechanism is the exchange of one virtual photon of mass $q^2=-Q^2$ (OPE), an elegant and simple formalism allows to express the electromagnetic current of a spin $S$ hadron in terms of $2S+1$ electromagnetic form factors (FFs).

In this contribution, we discuss the proton structure and the reaction mechanism for annihilation and scattering reactions in the energy range of a few GeV. Unpolarized electron proton elastic scattering has been considered the simplest way to determine FFs, using the Rosenbluth separation: measurements at fixed  
$Q^2$, for different angles (which requires to change the beam energy and the spectrometer setting) allow to 
extract the electric $G_E$ and the magnetic $G_M$ proton form factors, through the slope and the intercept of the Rosenbluth plot. 

In recent years it has been possible to measure the polarization of the outgoing proton, scattered by a longitudinally polarized beam. The ratio of the transverse to longitudinal polarization of the scattered proton $P_T/P_L$ allows to access the ratio of the electric to magnetic form factor, not only their squared values as the cross section does. This method, suggested by A.I. Akhiezer and M.P. Rekalo \cite{Re68}, is more sensitive to a small contribution of the electric FF, especially at large values of $Q^2$, where the magnetic contribution is dominant. The surprising result, which was obtained by the GEp collaboration \cite{GEp}, gave rise to a huge number of theoretical and phenomenological papers, and to a large experimental activity. Polarization experiments show that over $Q^2=1$ (GeV/c)$^2$
the $Q^2$ dipole approximation of FFs does not hold anymore and the FFs ratio follows a straight line: $ R=\mu G_E/G_M= 1.059 -0.143 ~Q^2~[(\mbox{GeV/c})^2]$ at least up to  a $Q^2\sim 6$ (GeV/c)$^2$. 
 
As no bias has been found in the experiments, the reason for the discrepancy is possibly related to radiative corrections (RC). In unpolarized experiments, RC can reach 40\% and high order radiative corrections are typically not included in the data analysis. No RC are applied to the polarization ratio, as they are assumed to be negligible, and indeed, first order RC cancel. The magnetic FF can be considered well known from cross section measurements, so the common interpretation of the present results is that the electric distribution in the proton is different from what previously assumed. Being $G_E$ related to the slope of the Rosenbluth plot, it has been noticed \cite{ETG05} that radiative corrections are largely responsible for this slope, in particular the C-odd corrections which are due to brehmstrahlung, more exacly, to the interference between electron and proton soft photon emission and two photon exchange (TPE).

TPE can not be calculated in model independent way when the target is a proton. Exact calculations for a lepton target, in a pure QED framework, have been done and show that TPE can not exceed 1-2\% \cite{Ku06}. Moreover, the lepton case is, by definition, an upper limit of the case where the target is a proton and the intermediate state in the box diagram is also a proton \cite{Ku08a}. TPE calculations for a proton target require modelization and different calculations have been done, with no quantitative agreement (see \cite{twof} and refs therein). Let us mention that in \cite{Ku08} it has been argued that elastic and inelastic intermediate states in the box diagram esentially cancel, due to analytical properties of the reaction amplitude.

Standard radiative corrections take into account 
the contribution of TPE where most of the transferred momentum 
is carried by one photon, while the other photon has very small momentum. The TPE contribution is larger when the two photons share equal momentum, and, due to the steep decrease of FFs, it might compensate the $\alpha$ counting rule. 
If such mechanism is important, it should be more visible at larger values of momentum transfer and for hadrons heavier than proton.
 
Crossing symmetry, which holds at the level of Born diagram, allows to relate 
the matrix elements ${\cal M}$ of the crossed 
processes, through the amplitude $f(s,t)$ :
\begin{equation}
\overline{|{\cal M}(eh\rightarrow eh)|^2}=f(s,t)=\overline{|{\cal 
M}(e^+e^-\rightarrow \overline h h)|^2}.
\end{equation}
The line over ${\cal M}$ denotes the sum over the polarizations of all 
particles (in initial and final states), $s$ and $t$ are the Mandelstam variables, which span different kinematical regions for annihilation and scattering channels.
 
The presence of a single virtual photon in the reaction 
$e^+e^-\rightarrow\gamma^*\rightarrow\overline{h}{h}$ constrains the
total angular momentum ${\cal J}$ and the $P$-parity 
for the $\overline{h}{h}-$system, to take only one 
possible value, ${\cal J}^P=1^-$, the quantum number of the photon. Therefore, 
in the framework of one-photon exchange (OPE), the 
$\cos \theta$-dependence of $\overline{|{\cal M}(eh\rightarrow eh)|^2}$ 
can be predicted in a general form ($\theta$ is the CMS angle between the momenta of the electron and the detected antinucleon):
\begin{equation}
\overline{|{\cal M}(eh\rightarrow eh)|^2}=a(t)+b(t)\cos^2\theta,
\label{eq:cse}
\end{equation}
where $a(t)$ and $b(t)$ are definite quadratic combinations of the 
electromagnetic form factors for the hadron $h$. The C-invariance of the electromagnetic hadron interaction allows only even powers of $\cos\theta$ and the degree of the $\cos\theta$-polynomial is limited to the second order, due to the spin of the virtual photon. One can show \cite{Re99} that there is a one-to-one correspondence between $\cos^2\theta$ and $\cot^2(\theta_e/2)$ ($\theta_e$ is the LAB angle of the emitted electron in the scattering channel) which explains 
the 
origin of the linear $\cot^2\frac{\theta_e}{2}$-dependence of the differential cross section for any $eh$-process in OPE approximation. 

The presence of TPE in the intermediate state: 
$e^++e^-\rightarrow 2\gamma\rightarrow\overline{h}+{h}$ can induce any value of the total angular momentum and space parity in the annihilation channel, but the $\overline{h}h$-system, produced through OPE and TPE mechanisms has different values of C-parity, 
because $C(\gamma)=-1$ and $C(2\gamma)=+1$. Therefore the interference contribution to the 
differential cross section in the annihilation channel must be an \textbf{odd} function of $\cos\theta$. 

These model independent statements allow to sign the presence of TPE: non linearities in the Rosenbluth plot in the scattering channel, and odd $\cos\theta$ contributions in the differential cross section for the annihilation channel. The illustration of different sets of data is given below, for $e^-+\!^4He$ elastic scattering, $e^++e^-\to \bar p+ p$ annihilation and $e^{\pm}+p$ elastic scattering.
\section{Scattering channel}

The search of model independent evidence of TPE in the experimental data which should appear as a non linearity of the Rosenbluth fit was firstly done in case of deuteron in Ref. \cite{Re99} and in case of proton in Ref. \cite{ETG05}. No evidence was found, in the limit of the precision of the data.

Let us note, that these experiments are sensitive to the real part of the interference between OPE and TPE. Very precise measurements of the transverse beam spin asymmetry in elastic electron scattering on proton and  $^4\!He$ suggest a non zero imaginary part of the TPE amplitude \cite{Ma04,K07}. Of particular interest is the case of $^4\!He$ target:
\be
e^-(p_1) +^4\! He(q_1)\to e^-(p_2)+^4\!He(q_2),
\label{eq:eq1}
\ee
as the spin structure of the matrix element is highly simplified for a spinless target. Using the
general properties of the electron--hadron interaction, such as the Lorentz
invariance and P--invariance, taking into account the identity of the initial
and final states and the T--invariance of the strong interaction, the scattering of a spin 1/2 particle on a spin zero target is described by two independent
amplitudes and the general form of the matrix element can be written independently from the reaction mechanism, as :
\be
{\cal M}(s,q^2) =\frac{e^2}{Q^2}\bar u(p_2)\Bigg [mF_1(s,q^2)+
F_2(s,q^2)\hat P\Bigg ]u(p_1)\varphi (q_1)\varphi (q_2)^*=\frac{e^2}{Q^2}{\cal N},
\label{eq:eqmat}
\ee
where $\varphi (q_1)$
and $\varphi (q_2)$ are the wave functions of the initial and final helium, with $P=q_1+q_2$ and $u(p_1)$, $u(p_2)$ are the spinors of the initial and final electrons, respectively. Here $F_1$ and $F_2$ are two invariant amplitudes, which are,
generally, complex functions of two variables $s=(q_1+p_1)^2$ and
$q^2=(q_2-q_1)^2=-Q^2$ and $m$ is the electron mass.  The matrix element (\ref{eq:eqmat}) contains the helicity--flip amplitude $F_1$ proportional to the electron mass which is 
explicitly singled out. The single--spin asymmetry, of interest here, is proportional to $F_1$. In OPE approximation one has:
\be
F_{1}^{Born}(s,q^2)=0, \  F_{2}^{Born}(s,q^2)=F(q^2),
\label{eq:eqampb}
\ee
where $F(q^2)$ is the helium electromagnetic charge form
factor depending only on $q^2$, with normalization $F(0)=Z,$ where $Z$ is the helium charge.

To separate the effects due to the Born and the 
two--photon exchange contributions, let us define the following decompositions of the amplitude \cite{Ga08a}
\be
F_{2}(s,q^2)=F(q^2)+f(s,q^2)\mbox{~where~}F_{1}(s,q^2) \sim\alpha,~f(s,q^2)\sim\alpha,\mbox{~and~}~F(q^2)\sim \alpha^0.
\label{eq:eqtwo}
\ee
Since the terms
$F_{1}$ and $f$ are small in comparison with the dominant one, one can safely neglect the bilinear combinations of these small terms multiplied by the factor $m^2$.

The differential cross section of the reaction (\ref{eq:eq1}), for the case of
unpolarized particles, has the following form in the Born approximation
\be
\frac{d\sigma_{un}^{Born}}{d\Omega}=\frac{\alpha ^2\cos^2\frac{\theta}{2}}
{4E^2\sin^4\frac{\theta}{2}}\Biggl [1+2\frac{E}{M}\sin^2\frac{\theta}{2}
\Biggr ]^{-1}F^2(q^2),
\label{eq:aqb}
\ee
where $\theta $ is the electron scattering angle in Lab system and $M$ is the helium mass. 

\begin{wrapfigure}[22]{R}{80mm}
  \centering 
\includegraphics[width=8cm]{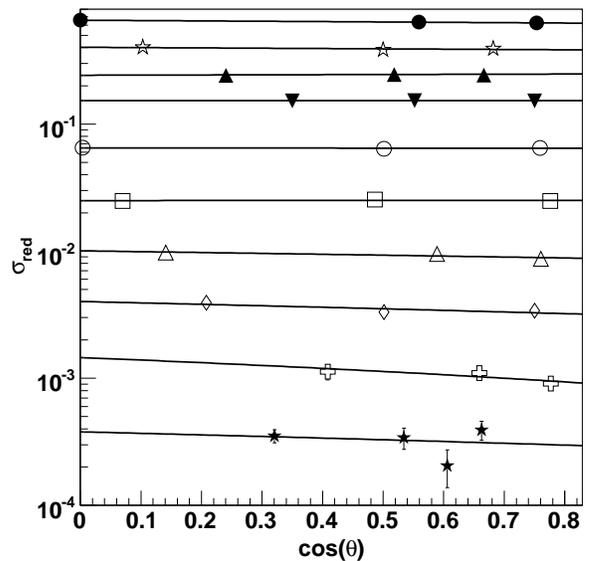}
\caption{\small Reduced cross section as a function of $\cos\theta$, at $Q^2$=0.5, 1, 1.5, 2, 3, 4, 5, 6, 7, 8 fm$^{-2}$ (from top to bottom). The data are from Ref. \protect\cite{Fr67}. Lines are two parameter linear fits.  }
\label{Fig:Ros}
\end{wrapfigure}

In the Born approximation, the $^4\!He$ FF depends only on the momentum transfer squared, $Q^2$. The presence of a sizable TPE contribution should appear as a deviation from a constant behavior of the reduced cross section measured at different angles and at the same $Q^2$. In case of $^4\!He$ few data exist at the same $\bar Q^2$ value, for $Q^2<$ 8 fm$^{-2}$ \cite{Fr67}. No deviation of these data from a constant value is seen, from a two parameter linear fit. The slope for each individual fit is always compatible with zero (see Fig. \ref{Fig:Ros}). 

The TPE contribution leads to new terms in the differential cross section :
\ba
\frac{d\sigma_{un}}{d\Omega}&=&\frac{\alpha ^2\cos^2\frac{\theta}{2}}
{4E^2\sin^4\frac{\theta}{2}}\Biggl [1+2\frac{E}{M}\sin^2\frac{\theta}{2}
\Biggr ]^{-1}\Biggl\{F^2(q^2)+2F(q^2)Re~f(s,q^2)+|f(s,q^2)|^2+
\nn \\
&&+\frac{m^2}{M^2}\Biggl [\frac{M}{E}+(1+\frac{M}{E})\tan^2\frac{\theta}{2} 
\Biggr ]F(q^2)ReF_1(s,q^2)\Biggr\},  
\label{eq:aqt}
\ea
and to a non--zero asymmetry,  in the case of the elastic scattering of 
transversally polarized electron beam. Let us define a coordinate frame with  $z \parallel p$, 
$y \parallel {\vec p}\times {\vec p}'$, where ${\vec p}({\vec p}')$ is the initial
(scattered) electron momentum, and the $x$ axis directed to form a left--handed
coordinate system. The transverse asymmetry, due to the 
interference between OPE and TPE, is determined by the polarization component perpendicular to the reaction plane:  
\be
A_y=\frac{\sigma^{\uparrow} -\sigma^{\downarrow}}
{\sigma^{\uparrow} +\sigma^{\downarrow}}\sim {\vec s}_e\cdot \frac{{\vec p}\times {\vec p}'}
{|{\vec p}\times {\vec p}'|}\equiv s_y,
\label{eq:aqy}
\ee
where $\sigma^{\uparrow} (\sigma^{\downarrow})$ is the cross section for electron beam
polarized parallel (antiparallel) to the normal of the scattering plane and ${\vec s}_e$ is the spin vector of the electron beam.
In terms of the amplitudes, it is expressed as:
\be
A_y=2\frac{m}{M}\tan\frac{\theta}{2}\frac{ImF_{1}(s,q^2)}{F(q^2)}.
\label{eq:eqasym}
\ee
Being a T--odd quantity, it is
completely determined by  the TPE contribution through the the imaginary part of the spin--flip amplitude $F_1(s,q^2)$ and, therefore, it is proportional to the electron mass.

For elastic  $e^-+^4\!He$ scattering, a value of $A_y^{exp}(^4\!He)=-13.51\pm 1.34(stat)
\pm 0.37(syst)$ ppm  for $E =2.75$ GeV, 
$\theta = 6^0$, and $Q^2$=0.077  GeV$^2$ has been measured \cite{K07}, to be compared to a theoretical prediction
$A_y^{th}(^4\!He)\approx 10^{-10}$ \cite{CH05}. The difference (by five orders of magnitude) was possibly  explained by a significant contribution of the excited states of the nucleus.

The measured value of the asymmetry allows to determine the size of the
imaginary part of the spin--flip amplitude $F_1$ \cite{K07}. From Eq. (\ref{eq:eqasym}) we  obtain $Im ~F_1\approx -F(q^2)$ for $\theta =6^0$. Assuming that $Re~F_1\approx Im~F_1$, then the contribution of $F_1$ to the differential 
cross section is negligible due to
the small factor $m^2/M^2$. One may expect that the imaginary part of the 
non--spin--flip amplitude, namely, its TPE part, is of the
same order as $Im~F_1$ since we singled out the small factor $m/M$ from the 
amplitude $F_1$. In this case we obtain an extremely large value for the 
TPE mechanism, of the same order as the 
OPE contribution itself, at such low $q^2$ value. We can conclude that either our assumption, about the magnitudes of $Im~f$ and $Im~F_1$, is not correct, or the experimental results on the asymmetry are somewhat large.

\section{Annihilation channel}

The general analysis of the polarization phenomena
in the reaction 
\be
e^+(p_+) +e^-(p_-)\to p(p_1)+\bar p(p_2) 
\label{eq:eqr}
\ee 
and in the time reversal channel, taking into account the TPE contribution, was done in Refs. \cite{Ga06}. 

The TPE contribution, if present, should also manifest itself in the time--like region. From first principles,  as the C-invariance of the electromagnetic interaction and the crossing symmetry, the presence of TPE would create a forward backward asymmetry in the differential angular distribution of the emitted particle. 
 
Such angular distributions have been recently measured by Babar \cite{Ba06}, for different ranges of the invariant mass of the $p\bar p$ pair, after selection of the reaction (\ref{eq:eqr}) by tagging a hard photon from initial state radiation (ISR). The distributions have been built with the help of a Monte Carlo (MC) simulation, which takes into account the properties of the detection and allows to subtract the background. 

\begin{wrapfigure}[15]{R}{70mm}
  \centering 
  \vspace*{-2mm} 
\includegraphics[width=7cm]{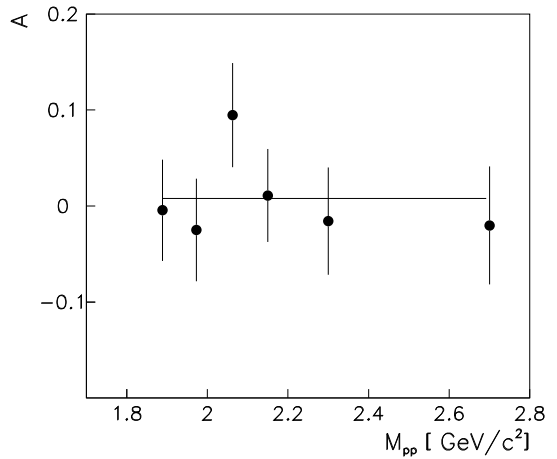}
\vspace*{-1.8truecm} 
\caption{\small Average forward backward asymmetry as a function of $M_{p\bar p}$.}
\label{Fig:fig3}
\end{wrapfigure}

For each $\cos\theta$ bin and each invariant mass interval, the angular asymmetry is defined as:
\be
{\cal A}(c)=\displaystyle\frac{
\displaystyle\frac{d\sigma}{d\Omega}(c) -
\displaystyle\frac{d\sigma}{d\Omega}(-c)}
{\displaystyle\frac{d\sigma}{d\Omega}(c) +\displaystyle\frac{d\sigma}{d\Omega}(-c)},~c=\cos\theta \mbox{~and~} {\cal A}(0)=0.
\label{eq:eqas}
\ee
The dependence of the asymmetry as a function of $\cos\theta$ is rather flat and can be fitted by a constant, in each mass range. The experimental values of the asymmetry are compatible with zero for all mass ranges, with a typical error is $\sim 5\%$. As no systematic effect over $M_{p\bar p}$ appears, one can calculate the global average: ${\cal A}=0.01 \pm 0.02$ (Fig. \ref{Fig:fig3}).

Note that radiative corrections of C-odd nature could also contribute to an eventual asymmetry in the data. Other odd contributions to the reaction (\ref{eq:eqr}), with respect to  $\cos\theta$, may arise due to $Z$-boson exchange and C-odd interference of radiative amplitudes (including the emission of virtual and real photons). For energies smaller than the $Z$-boson mass, $\sqrt{t}/M_Z\ll 1$, the $Z-$boson exchange can be neglected. 

The largest contribution to the asymmetry is represented by a factor which depends on the soft photon energy $\Delta E$ and is partially compensated by hard photon emission, with energy $\omega>\Delta E$. The hard contribution to the asymmetry $A^{hard}$ was explicitely calculated in Ref. \cite{Ku77}.  
The total contribution to the asymmetry is expected to be $A^{tot}=A^{soft}+A^{hard}\le 2\%$ \cite{ETG08}. 

The total contribution from radiative corrections to the angular asymmetry is not expected to exceed 2\%. Moreover, radiative corrections have been applied to the data, therefore part or all of the asymmetry arising by soft and hard photon emission is already taken into account in the differential cross section. 

The analysis of the available data shows no asymmetry, within an error of 2\%. Such error is of the order of the asymmetry expected from radiative corrections as calculated from QED. As no systematic deviations are seen, we can conclude that these data do not give any hint of the presence of TPE, in all the considered kinematical range.

\section{Electron and positron scattering }

In the Born approximation, the elastic cross section is identical for positrons and electrons. A deviation of the ratio: 
\be 
R=
\displaystyle\frac{\sigma (e^+ h\to e^+h )}
{\sigma (e^- h\to e^-h )}=\frac{1+A^{odd}}{1-A^{odd}}
\label{eq:ratio}
\ee
from unity would be a clear signature of processes beyond the Born approximation. Those processes include the interference of OPE and TPE, and all the photon emissions which bring a C-odd contribution to the cross section.

A model for TPE in $e^{\pm} p$ scattering was derived in Ref. \cite{Ku08}, and the charge asymmetry:
\ba
A^{odd}&=&
\frac{d\sigma^{e+p}-d\sigma^{e^-p}}{d\sigma^{e+p}+d\sigma^{e^-p}}
 =-\frac{2\alpha }{\pi}
\left [ \ln\frac{1}{\rho}
\ln\frac{Q^2 x}{2\rho (\Delta E)^2}+
\Li\left ( 1-\frac{1}{\rho x} \right )-
\right .\nn \\
&&
\left .
\Li\left( 1-\frac{\rho}{x}\right )
\right ], ~
~x=\frac{\sqrt{1+\tau}+\sqrt{\tau}}{\sqrt{1+\tau}-\sqrt{\tau}},~\tau=\frac{Q^2}{4M^2}
\label{eq:eqv1}
\ea
was expressed as the sum of the contribution of two virtual photon exchange, (more exactly the interference between the Born amplitude and the box-type amplitude) and a term which depends on the maximum energy of the soft photon, which escapes the detection, $\Delta E$ ($E$ is the initial energy and $\rho$ is the fraction of the initial energy carried by the scattered electron). It turns out that it is namely this term which  gives the largest contribution to the asymmetry and contains a large $\epsilon$ dependence ($\epsilon^{-1}=1+2(1+\tau)\tan^2(\theta_e/2)$. Note that Eq. (\ref{eq:eqv1}) holds at first order in $\alpha$ and does not include multi-photon emission.
\begin{wrapfigure}[21]{R}{80mm}
  \centering 
  \vspace*{-8mm} 
\includegraphics[width=80mm]{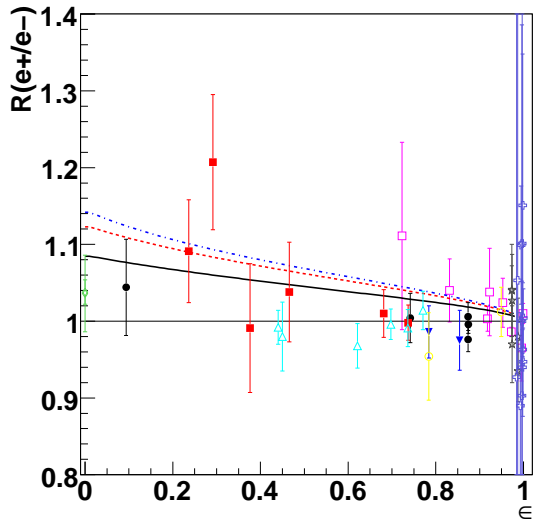}
\vspace*{-0.8truecm}
\caption{\small Ratio of cross sections $R=\sigma(e^+p)/\sigma(e^-p)$, as a function of $\epsilon$, for $c=0.97$ and $Q^2=1$ GeV$^2$ (solid line, black) $Q^2=3$ GeV$^2$ (dotted line, red) and  $Q^2=5$ GeV$^2$ (dash-dotted line, blue). }
 \label{fig:epem_R_eps}
\end{wrapfigure}

Let us note that a C-odd effect is enhanced in the ratio (\ref{eq:ratio}) with respect to the asymmetry (\ref{eq:eqv1}). Experiments on elastic and inelastic scattering of $e^+$ and $e^-$ beams in identical kinematical conditions  have been performed and recently reviewed in \cite{ETG09}.

The elastic data are shown in Fig. \ref{fig:epem_R_eps} as a function of $\epsilon$ and compared to the model of Ref. \cite{Ku08} plotted at three values of $Q^2$. The comparison of the calculation with the data has to be performed point by point for the corresponding $(\epsilon,Q^2)$ values. The agreement turns out to be very good, in the limit of the experimental errors, showing that a possible deviation of this ratio from unity is related to soft photon emission. One can conclude that the data on the cross section ratio are compatible with the assumption that the hard two-photon contribution is negligible. 

\section{Conclusions}

In the scattering and annihilation channels involving the electron proton interaction, the presence of TPE can be parametrized in a model independent way. In the scattering channel, the additional terms induced by TPE depend on the angle of the emitted particle, and manifest as an angular dependence of the reduced  differential cross section at fixed $Q^2$.
The TPE contribution could also be detected using a transversally polarized electron beam, through a T-odd asymmetry of the order of the electron mass. An analysis of the existing data does not allow to reach evidence of the presence of the TPE mechanism for $^4\!He$, as well as for other reactions involving protons and deuterons. 

In the annihilation channel, the analysis of the BABAR data on $e^++e^-\to \bar p +p$,  in terms of $ \cos\theta$ asymmetry of the angular distribution of the emitted proton, does not show evidence of TPE, in the limit of the uncertainty of the data. 

The difference in the cross section for $e^{\pm} p$ scattering can be explained by odd terms, which are present in standard radiative corrections.

One can conclude that the data do not show evidence for the presence of the TPE at the level of their precision. TPE is expected to become larger when the momentum transfer increases. Its study in the kinematical range covered by the present experiments requires more precise and dedicated measurements. In this respect, future antiproton beams at the FAIR facility in Darmstadt will provide very good conditions for the measurement of time-like proton form factors and for the study of the reaction mechanism.

The work presented here was initiated in collaboration with Prof. M.P. Rekalo. These results would not be have been obtained without the collaboration of G.I. Gakh, E.A. Kuraev, S.~Bakmaev, V.V. Bytev, Yu. M. Bystritskiy, S. Pacetti and M.~Osipenko.

\end{document}